\documentclass{JHEP3}
\title{The conformal anomaly of  k-strings }
\author{Pietro Giudice, Ferdinando Gliozzi, Stefano Lottini\\
Dipartimento di Fisica Teorica, Universit\`a di Torino and \\
INFN, Sezione di Torino\\
via P.Giuria 1, I-10125 Torino, Italy\\
\email{giudice,gliozzi,lottini@to.infn.it}}
\abstract{Simple scaling properties of correlation functions of a
confining gauge theory in d-dimensions lead to the conclusion that 
k-string dynamics is described, in the infrared limit, by a two-dimensional 
conformal field theory with conformal anomaly $c=(d-2)\sigma_k/\sigma$, where $\sigma_k$ is the k-string  tension and $\sigma$ that of the 
fundamental representation. This result applies to any gauge theory with stable k-strings. We check it in a 3D $\Z_4$ gauge 
model at finite temperature, where a string effect directly related to $c$ 
can be clearly identified.}
\keywords{Lattice Gauge Field Theories, Confinement}
\usepackage[dvips]{graphicx}
\usepackage{amsmath}
\usepackage{amsfonts}
\usepackage{mathrsfs}
\usepackage{epsfig}
\usepackage{psfrag}
\newcommand{\Z}{\mathbb{Z}}
\newcommand{\A}{\mathbb{A}}

\newcommand{\R}{\mathscr{R}}

\newcommand{\U}{{\cal U}}
\newcommand{\Q}{{\cal Q}}
\newcommand{\bra}{\langle}
\newcommand{\ket}{\rangle}
\newcommand{\avg}[1]{\langle \hspace{0.2em} #1 \hspace{0.2em} \rangle}

\newcommand{\sun}{\mathop{\rm SU}(N)}

\newcommand{\eq}{\begin{equation}}
\newcommand{\en}{\end{equation}}
\newcommand{\bea}{\begin{eqnarray}}
\newcommand{\ea}{\end{eqnarray}}

\newcommand{\link}[1]{\bra #1\ket}

\begin{document}
\section{Introduction}

One of the simplest and most general consequences of the effective string 
description of the quark confinement \cite{first,Nambu:1974zg,Nambu:1978bd}
is the presence of measurable effects on physical observables of the 
gauge theory, produced by the quantum fluctuations of that 
string \cite{lsw,lu}. The most widely known is the L\"uscher  
correction to the interquark confining potential $V$ at 
large distance $r$
\eq
V(r)=\sigma\,r+2\,\mu-(d-2)\frac\pi{24r}+O(1/r^2)\,,
\label{pot}
\en
where $\sigma$ is the string tension  and $\mu$ a self-energy term. The 
attractive, Coulomb-like correction is universal in the sense that it is 
the same for whatever confining gauge theory and depends only on the $d-2$ 
transverse oscillation modes of the   string.

A similar universal effect has been found in the low temperature behaviour 
of the string tension \cite{pa}
\eq
\sigma(T)=\sigma-(d-2)\frac\pi6 T^2+O(T^4)\,.
\label{st}
\en
Both effects may be rephrased by saying that the infrared limit of the 
effective string is described by a two-dimensional conformal field 
theory (CFT) with conformal anomaly $c=d-2$. In this language, the 
(generalised) L\"uscher term $-\frac{c\,\pi}{24\,r}$ is the zero-point energy 
of a 2D system of size $r$ with Dirichlet boundary conditions, while the 
$-\frac{c\,\pi}{6}T^2$ term 
is the zero-point energy density in a 
cylinder (i.e. the string world-sheet of the Polyakov correlator) of period 
$L=1/T$ \cite{bcn}. 

In this paper we propose a method to extend these 
results to a more general class of confining objects of $\sun$ gauge theories, 
the k-strings, describing the infrared behaviour of the flux tube joining 
sources in representations with $N-$ality $k$.      
There are of course infinitely many irreducible 
representations corresponding to the same value of $k$. No matter what 
representation $\R$ is chosen, the stable string tension $\sigma_k$  
depends only on the $N-$ality $k$ of $\R$, i.e. on the number (modulo $N$) 
of copies of the fundamental representation needed to build $\R$ by tensor 
product, since the sources 
may be screened by gluons. As a consequence, at 
sufficiently large 
distances, the heavier strings find it energetically favourable to decay 
into the string of smallest string tension, called k-string.

The spectrum of k-string tensions has been extensively studied in 
recent years, in the continuum \cite{ds}--\cite{jr} as well  as           
on the lattice \cite{lt1}--\cite{ddpv}. 
So far, in numerical analyses one typically measured  the 
temperature-dependent k-string tensions $\sigma_k(T)$ through 
the Polyakov correlators and then extrapolated to $T=0$ using (\ref{st}), 
hence assuming  a  free bosonic string behaviour. 

Recently this assumption has been questioned 
by a numerical experiment. It showed that in a 3D $\Z_4$ gauge theory, 
though the 1-string fitted  perfectly the free string formulae with 
a much higher precision than in the $\sun$ case, the 2-string failed to 
meet free string expectations \cite{z4kstr_pietro}.
One could object that there is no compelling reason for a 2-string of a 
$\Z_4$ gauge system to  behave like a 2-string of  $\sun$ gauge system; 
a k-string can be seen as a bound state of $k$ 1-strings and the binding 
force would presumably depend on the specific properties of the gauge system.
  
On the other hand, from a theoretical point of view there are good reasons
to expect values of $c$ larger than $d-2$. In fact the conformal anomaly 
can be thought of as counting the number of degrees of freedom of the 
k-string. Therefore the relevant degrees of freedom are not only the 
transverse displacements but also the splitting of the k-string into its 
constituent strings. If the mutual interactions where negligible,  
each constituent string could vibrate independently so we had 
$c=k\,(d-2)$. Thus we expect that $c$ can vary in the range
\eq
d-2 \le\, c\,\le k\,(d-2)\,. 
\en

An unexpected insight into the actual value of $c$ comes when 
considering the infrared properties of the N-point Polyakov correlators related to 
the baryon vertex of $\sun$
\eq
\bra\, P_f(\vec{r}_1)\,P_f(\vec{r}_2)\dots P_f(\vec{r}_N)\,\ket_T\,,
\en
where $P_f(\vec{r}_i)$ is the Polyakov line in the fundamental representation  
identified by  $d-1$ spatial coordinates $\vec{r}_i$  and directed along 
the periodic temporal direction of size $L=1/T$. If all the distances 
$\vert\vec{r}_i-\vec{r}_j\vert$ are much larger than any other relevant scale, 
this correlator is expected to obey a simple scaling property. 
When combining this fact 
with the circumstance that, depending on the positions $\vec{r}_i$ of the 
sources, some strings of the baryon vertex may coalesce into k-strings 
\cite{Hartnoll:2004yr,bary}, one obtains the geometric constraint
\eq
\sigma_k(T)=\sigma_k-(d-2)\frac{\pi\,\sigma_k}{6\,\sigma}\,T^2+O(T^3)\,,
\label{main}
\en 
which is the main result of this paper. We check this formula in a 
3D $\Z_4$ gauge model where, thanks to duality, very  efficient simulation
techniques are available, yielding high precision results which give fairly convincing evidence for the scaling law (\ref{main}). 

The contents of this paper are as follows. In the next Section we expose in 
detail the above-mentioned scaling argument, while in the following Section we
describe a lattice calculation on a three-dimensional $\Z_4$ gauge theory where
combining duality transformation with highly efficient simulation techniques 
it is possible to confirm that the stable 2-string matches nicely 
Eq.(\ref{main}). We finish with a discussion of our results and 
some of their implications.  

\section{Scaling form of the Polyakov correlators}
The main role of the string picture of confinement is to fix the functional form of the vacuum expectation value of gauge invariant operators in the infrared 
limit. It predicts two different 
asymptotic behaviours of the correlation function of a pair of Polyakov loops 
$\bra P_f(\vec{r}_1)\,P^\dagger_f(\vec{r}_2\ket_T$, when both 
$r\equiv\vert\vec{r}_1-\vec{r}_1\vert$ and $L\equiv1/T$ are large, 
depending on the value of the ratio $2r/L$. Using (\ref{pot}) and 
(\ref{st}) we can write
\eq
-\frac{1}{L}\log\bra P_f(\vec{r}_1)\,P^\dagger_f(\vec{r}_2)\ket_T=
 V(r)+ O(e^{-\pi\,L/r})~,~~2\,r<\,L\,,
\en
\eq
-\frac{1}{L}\log\bra P_f(\vec{r}_1)\,P^\dagger_f(\vec{r}_2)\ket_T=
 \sigma(T)\,r+2\mu+\frac{1}{2L}\log{\frac{2r}{L}}+O(e^{-4\pi r/L})~,~~2\,r>\,L\,,
\label{logppst}
\en
\noindent where $V(r)$ and $\sigma(T)$ are given by (\ref{pot}) 
and (\ref{st}). 
There are strong indications that the $r^{-3}$ and $T^3$  terms are zero  
and  the $r^{-4}$ and $T^4$ are universal (see for instance the discussion
in \cite{jk} and references quoted therein). For our purposes 
we need  only the first universal term, which is directly related to the 
central charge of the CFT describing the IR limit of the 
underlying confining string.
In this approximation the Polyakov loop correlation 
functions should decay at large $r$ while keeping constant $T$ as
\eq
\bra P_f(\vec{r}_1)\,P^\dagger_f(\vec{r}_2)\ket_T \stackrel{\sim}{\propto}
\exp\left[-\sigma(T)\,r\,L-2\mu\,L\right]\,.
\label{ppst}
\en
Similar expansions are expected to be valid also for Polyakov correlators 
describing more specific features of $\sun$ gauge theory, like 
those involving baryonic vertices.
\FIGURE{\epsfig{file=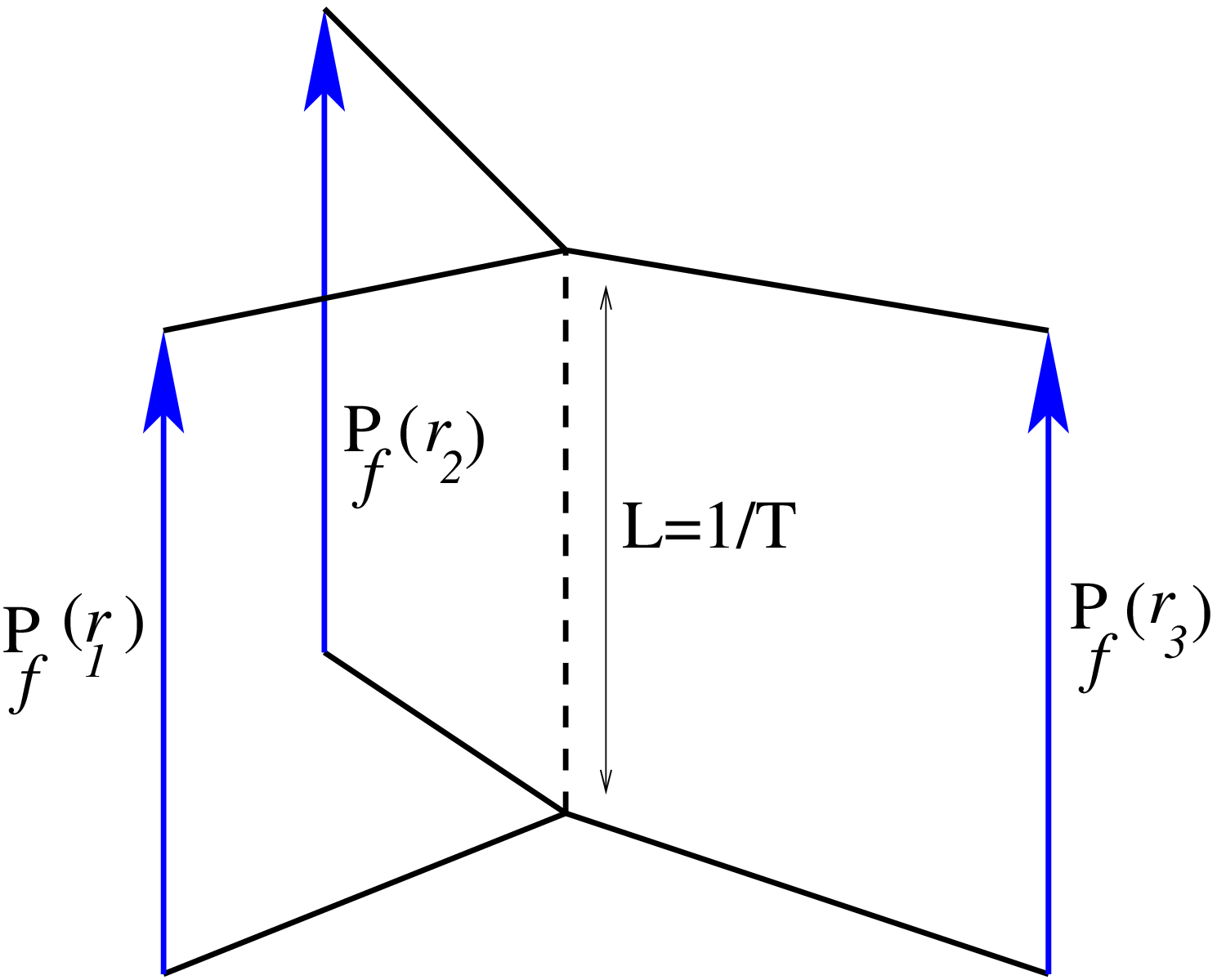,width=7.5 cm}
\caption{The three-bladed world-sheet of a static SU(3) baryon. 
\label{Figure:1}}}
A baryon vertex is a gauge-invariant coupling of $N$ multiplets in the 
fundamental representation $f$ which gives rise to
configurations of finite energy, or baryonic potential, with $N$ 
external sources. 
At finite temperature $T$ these 
sources are the Polyakov lines $P_f(\vec{r}_i)$.
Assume for a moment $N=3$. When the mutual distances 
$\vert\vec{r}_i-\vec{r}_j\vert$  $(i,j=1,2,3)$ are all large,  
three confining strings of chromo-electric flux form, which, starting from 
the three sources,  meet a common junction; their world-sheet forms a 
three-bladed surface with a common intersection 
\cite{deForcrand:2005vv,Bissey:2006bz} (see Fig.\ref{Figure:1}).    

The total area $\A$ of this world-sheet is
 $L\,\ell(\vec{r}_1,\vec{r}_2 ,\vec{r}_3 )$ where $\ell$ is the total 
length of the string. The stable configuration (hence the position of the 
common junction) is the one minimising $\A$. The balance of tensions implies 
angles of $\frac{2\pi}3$ between the blades.

The complete functional form of the 3-point correlator is substantially 
unknown. Nonetheless, the static baryon potential, defined as 
\eq
V=-\lim_{T\to\infty}T\,\log\bra\, 
P_f(\vec{r}_1)\,P_f(\vec{r}_2)\, P_f(\vec{r}_3)\,\ket_T\,,
\en
has a simple form in  the IR limit:  
\eq
V=\sigma\,\ell(\vec{r}_1,\vec{r}_2 ,\vec{r}_3)+3\mu+\dots
\en
The universal $1/r$ corrections have been calculated in \cite{Jahn:2003uz}.

In the IR limit at finite temperature, i.e. 
$\vert\vec{r}_i-\vec{r}_j\vert\gg L ~ \forall i\not=j$,
we assume, in analogy with (\ref{ppst}),
\eq
\bra\, P_f(\vec{r}_1)\,P_f(\vec{r}_2)\dots P_f(\vec{r}_N)\,\ket_T
=e^{-F_N}
\sim\exp\left[-\sigma(T)\,L\,\ell(\vec{r}_1,\vec{r}_2 ,\dots,\vec{r}_N)
-N\mu\,L\right]\,,
\label{pnst}
\en
or, more explicitly,
\eq
F_N(\ell,L)\sim\sigma\,L\ell-(d-2)\frac{\pi\,\ell}{6\,L}+N\mu\,L~, \,
~(\vert\vec{r}_i-\vec{r}_j\vert\gg L ~ \forall i\not=j)\,,
\label{fb}
\en
where the coefficient of the $\ell/L$ term specifies that in this 
IR limit the behaviour of the baryon flux distribution is described by a 
CFT with conformal anomaly $d-2$  on the string world-sheet singled out 
by the position of the external sources. 

The above considerations are completely standard and  nothing 
new happens so far. 
 \FIGURE{
\includegraphics[width=7cm]{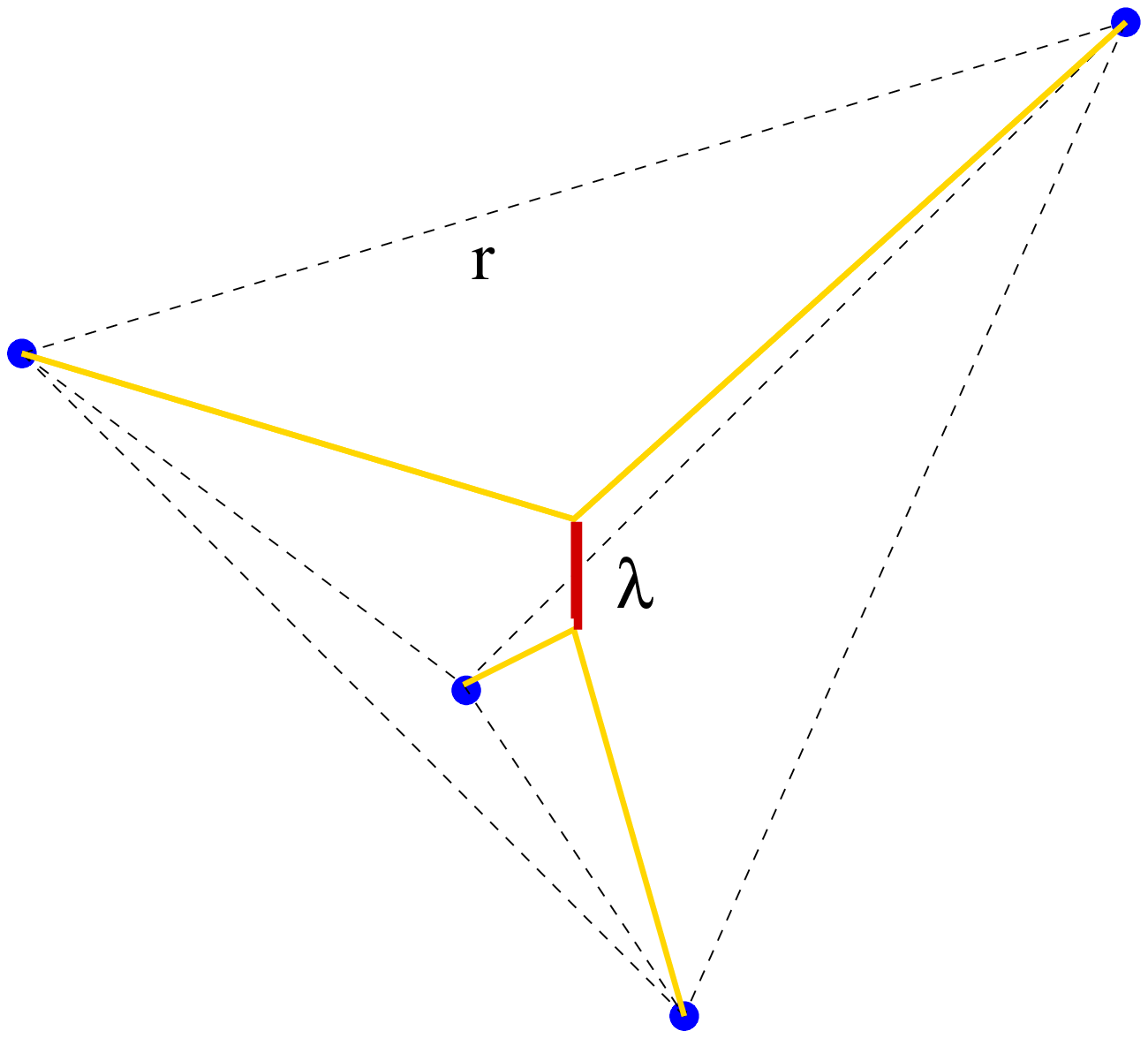}
\caption{Confining string configuration of a SU(4) static baryon. 
The thick line is a 2-string. \label{Figure:2}}}
\noindent The surprise comes when 
considering the latter expression in the case
 $N>3$. Notice that, depending on the positions $\vec{r}_i$ of the 
sources, some fundamental  strings of the baryon vertex may coalesce into 
k-strings \cite{Hartnoll:2004yr,bary}. As a consequence, the shape of the 
world-sheet changes in order to balance the string tensions and $\ell$ becomes 
a weighted sum, where a k-string of length $\lambda$ contributes with a term
$\lambda\,\frac{\sigma_k(T)}{\sigma(T)}\,$, where $\sigma(T)$ is given by 
(\ref{st}), while
\eq
\sigma_k(T)=\sigma_k-c_k\,\frac\pi6 T^2+O(T^3)\,,
\label{skt}
\en
where $c_k$ is the conformal anomaly of the k-string. 
When the string is in the fundamental representation the $T^3$ term 
(\ref{st}) is missing 
\cite{lw4,dr,hdm}, however we cannot exclude it in the k-string
with $k>1$. The string tension ratio is
\eq
\frac{\sigma_k(T)}{\sigma(T)}=\frac{\sigma_k}{\sigma}
+\frac{\sigma_k}{\sigma} \left(\frac{d-2}\sigma-\frac{c_k}{\sigma_k}\right)
\frac\pi6 T^2+O(T^3)\,;
\label{rskt}
\en
if the coefficient of the $T^2$ term does not vanish the asymptotic 
functional form (\ref{fb}) of the free energy  gets modified.

As a simple, illustrative example, let us consider the chromo-electric 
flux distribution of a 4D $SU(4)$ gauge system generated by four external 
quarks placed at the vertices of a regular tetrahedron 
(see Fig.~\ref{Figure:2}).
Preparing the external sources in a symmetric configuration does not 
necessarily imply that the distribution of the gauge flux preserves 
the tetrahedral symmetry. In fact, the 
formation of a 2-string breaks this symmetry (see 
 Fig.\ref{Figure:2}). The actual symmetry breaking or restoration 
depends on the cost in free energy 
of the configuration, of course. It is easy to show that, when 
$\frac{\sigma_2}{\sigma}<\frac2{\sqrt{3}}$, the tetrahedral symmetry is 
spontaneously broken \cite{bary} and the length $\lambda$ of the 2-string, which is a function of the ratio $\sigma_2/\sigma$, is given by 
\eq
\lambda=\frac r{\sqrt{2}}-\frac{r\,\sigma_2}{\sqrt{4\,\sigma^2-
\sigma_2^2}}\,,
\label{la}
\en
where $r$ is the edge length. The free energy $F_4$ of the 
four-quark system is
\eq
F_4=\sigma(T)\,L\,\ell(T)+4\mu L=\sigma(T)\,L\,\left(2\,r\sqrt{1-
\left(\frac{\sigma_2(T)}{2\sigma(T)}\right)^2}+\frac{r}{\sqrt{2}}\,
\frac{\sigma_2(T)}{\sigma(T)}\right)+4\mu\, L\,.
\en    
Now the total length $\ell$ of the string may depend on $T$ through
the ratio $\sigma_2(T)/\sigma(T)$. Expanding in $T=1/L$ as in
(\ref{rskt})  we get
\eq
F_4=\sigma\,L\,\ell(0)-\frac{\pi}6\left[(d-2)\frac{\ell(0)}L-
{\sigma_2}\left(\frac{d-2}\sigma-\frac{c_2}{\sigma_2}\right)
\frac\lambda{L}\right]
+4\mu\,L+O(1/L^2)\,.
\en
Clearly the term proportional to $\lambda$ violates the expected 
asymptotic form of the free energy (\ref{fb}), unless 
$c_2=(d-2)\,\sigma_2/\sigma\,$.
\FIGURE{\epsfig{file=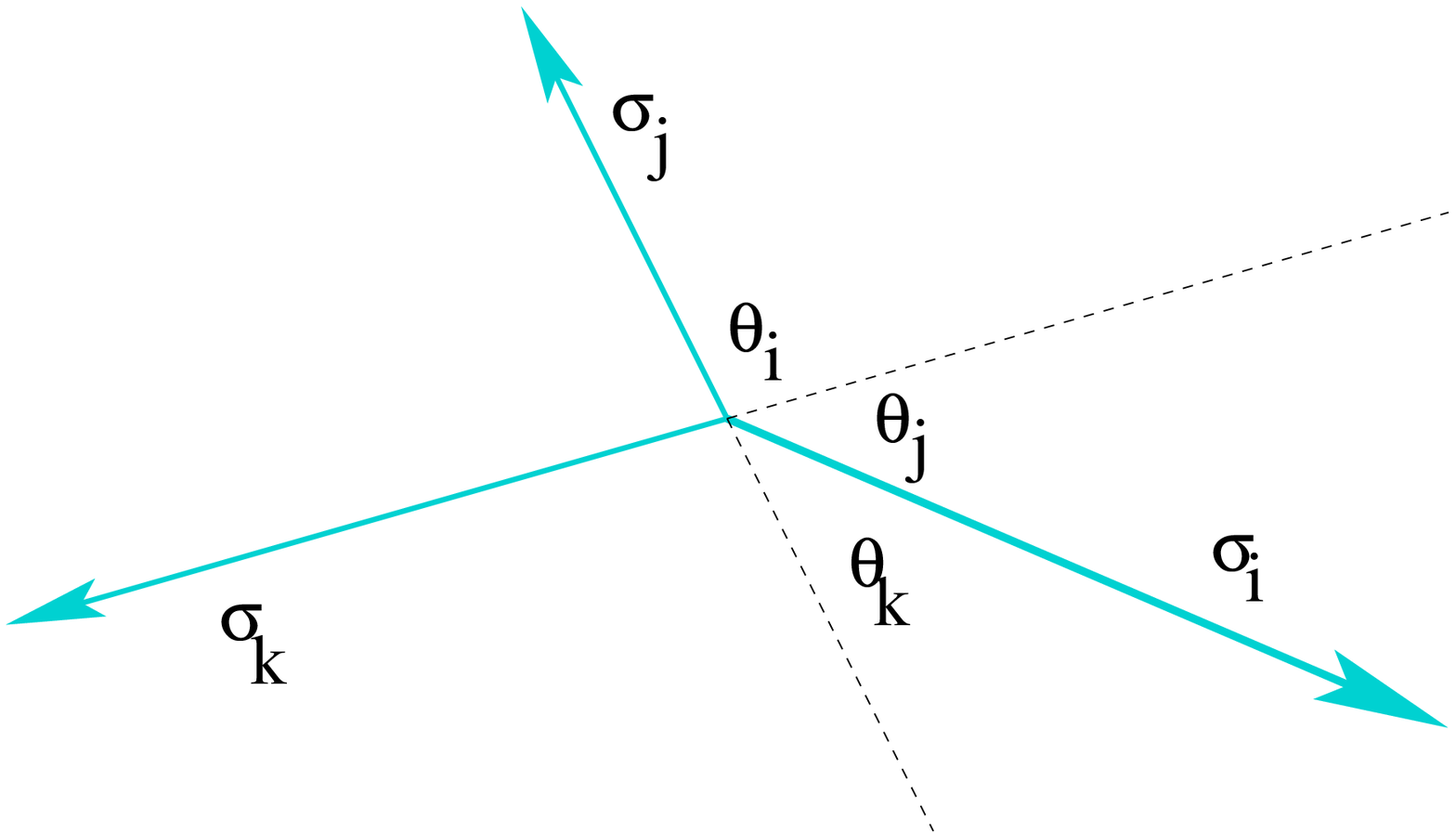,width=7 cm}\caption{The balance of the string tensions.
\label{Figure:3}}}
More generally, the baryonic free energy keeps the expected asymptotic 
form (\ref{fb}) only if the world-sheet shape does not change while varying 
$T$. Now  in a generic string configuration contributing to $\sun$ baryon  
potential, the angles at a junction of three arbitrary k-strings 
are given by (see Fig.\ref{Figure:3})
\eq
\cos\theta_i=\frac{\sigma_j(T)^2+\sigma_k(T)^2-\sigma_i(T)^2}
{2\,\sigma_j(T)\,\sigma_k(T)}
\en
and  others obtained by cyclic permutations of the indices 
$i,j,k$. As a consequence, these angles are kept fixed only if 
all the string tension ratios are constant up to $T^3$ terms, i.e. only if
\eq
\frac{c_i}{\sigma_i}=\frac{c_j}{\sigma_j}=\frac{c_k}{\sigma_k}=
\frac{(d-2)}{\sigma}
\en 
which leads directly to 
\eq
\sigma_k(T)=\sigma_k-(d-2)\frac{\pi\,\sigma_k}{6\,\sigma}\,T^2+O(T^3)
\en 
as anticipated in the Introduction.

\section{The 3D $\Z_4$ gauge model and its dual}

The above general argument on the finite temperature corrections 
of the k-string tensions suggests a different behaviour with 
respect to the usual assumption that these corrections are those 
produced by a free bosonic string.
Since the comparison with theoretical predictions of k-string 
tensions is sensitive to this behaviour, it is important to 
check its validity.

In this Section we address such a question with a lattice 
calculation in a 3D $\Z_4$ gauge theory which is perhaps 
the simplest gauge system where there is a stable 2-string. 

We work on a periodic cubic lattice $L\times L\times L_\tau$.
The degrees of freedom are the fourth roots of the identity
$\zeta_l$ $~(\zeta^4=1)$, defined on the links $l$ of the lattice.
The partition function is
\eq
Z(\beta_f,\beta_{ff})=\prod_l\sum_{\zeta_l=\pm1,\pm i}e^{\sum_p
(\beta_f\U_p+\beta_{ff}\U^2_p/2+c.c.)}\,,
\en
where the sum is extended to all plaquettes $p$ of the 
lattice and $\U_p=\prod_{l\in p}\zeta_l$; $\beta_f$ and $\beta_{ff}$
are two coupling constants. When they vary in a suitable range 
 the system belongs to a confining phase. In analogy with the 
$\sun$ case we say that $\U_p$ is in the fundamental ($f$)  
representation while $\U_p^2$ lies in the double-fundamental ($ff$) 
representation.
From a computational point of view it is convenient to  recast 
$Z$ as the partition function of two coupled 
$\Z_2$ gauge systems
\eq
	Z(\beta_f,\beta_{ff}) =\prod_l \sum_{ \{ U_l = \pm 1 ,
 V_l = \pm 1 \} }  e^{\sum_p\beta_f(U_p+V_p)+\beta_{ff} U_p V_p}\,,
~\left(U_p=\prod_{l\in p}U_l\,,V_p=\prod_{l\in p}V_l\right)\,.
\label{cz2}
\en
The external sources generating the 1-string 
and the 2-string are given respectively by the two products
\eq
P_f(\vec{r})\equiv\prod_{l\in\gamma_{\vec{r}}}U_l~{\rm or}~
\prod_{l\in\gamma_{\vec{r}}}V_l~;~~
P_{ff}(\vec{r})\equiv \prod_{l\in\gamma_{\vec{r}}}U_l\,V_l\,,
\label{pfpff}
\en 
where $\gamma_{\vec{r}}$ is a closed path in the lattice which 
winds once around the temporal direction $L_\tau$ and passes 
through $\vec{r}$.

This model, as any three-dimensional abelian 
gauge system, admits a spin model as its dual.  
We have recently shown \cite{Giudice:2006hw} that this $\Z_4$ gauge model 
is dual to a spin model with global $\Z_4$ symmetry which can 
be written as a symmetric Ashkin-Teller (AT) model, i.e.~two 
coupled Ising models defined by the two-parameter action
	\eq
	S_{AT} = -\sum_{\link{xy}} 
\left[ \beta (\sigma_x \sigma_y + \tau_x \tau_y) 
                 + \alpha (\sigma_x \sigma_y \tau_x \tau_y) \right]\,,
	\en
where $\sigma_x$ and $\tau_x$ are  Ising variables ($\sigma,\tau=\pm 1$) 
associated to the site $x$ and the sum is over all the links $\link{xy}$ of 
the dual cubic lattice. The two couplings $\alpha$ and $\beta$ are related to 
the gauge couplings $\beta_f$ and $\beta_{ff}$ by \cite{Giudice:2006hw} 

	\begin{eqnarray}
		\alpha & = &
	 \frac{1}{4} \log \left( \frac{(\coth\beta_f+\tanh\beta_f\tanh\beta_{ff})
(\coth\beta_f+\tanh\beta_f\coth\beta_{ff})}
                    {2+\tanh\beta_{ff}+\coth\beta_{ff}} \right) \,, \\
		\beta & = &
	 \frac{1}{4} \log \left( \frac{1+\tanh^2\beta_f\tanh\beta_{ff}}
{\tanh^2\beta_f+\tanh\beta_{ff}} \right) \, .
	\end{eqnarray}
The duality transformation maps any physical observable of the gauge theory  
into a corresponding observable of the spin model. In particular it is well 
known that the Wilson loops are related to suitable \emph{flips} of the 
couplings of the spin model.
We have found the identity
\eq
	\avg{V_P}_{gauge} = \avg{e^{-2(\beta+\alpha\tau_x\tau_y)
\sigma_x\sigma_y}}_{AT}
\en
generalising the known dual identity of the Ising model. Similarly, 
flipping the signs of both spins $\sigma_x$ and $\tau_x$ we get the
 plaquette variable in the $k=2$ representation as $\avg{U_P V_P}$. 
Combining together a suitable set of plaquettes we may build up any
 Wilson loop or Polyakov-Polyakov correlator with $k=1$ or $k=2$.

\subsection{Monte Carlo simulations}
Our interest in writing this model in terms of Ising 
variables is twofold.

 First we can easily apply for the simulation a very efficient non-local  
algorithm \cite{wd}, basically similar to the standard 
Fortuin-Kasteleyn cluster 
method of Ising systems: each update step is composed by an update of 
the $\sigma$ variables using the current values of the $\tau$'s as a 
background  (thus locally changing the coupling from $\beta$ to 
$\beta \pm \alpha$ according to the value of $\tau_x \tau_y$ on the link 
$\link{xy}$), followed by an update of the $\tau$'s using the $\sigma$ values 
as background.
\TABULAR{cc|c|c|}{\cline{3-4}
&&$\alpha=0.05$&$\alpha=0.07$\\
&&$\beta=0.2070$&$\beta=0.1975$\\
\hline
\multicolumn{2}{|c|}{$\sigma\equiv\sigma_f$}&0.02085(10)&0.0157(1)\\
\multicolumn{2}{|c|}{
$\sigma_2\equiv\sigma_{ff}$}&0.0328(5)&0.0210(5)\\
\multicolumn{2}{|c|}{$\sigma_2-\sigma$}&0.01195(51)&0.0053(5) \\
\hline
\multicolumn{1}{|c|}{~}&$N_\tau=9$& 0.00864(6)&--\\
\multicolumn{1}{|c|}{~}&$N_\tau=10$& 0.00951(8)&0.003700(30)\\
\multicolumn{1}{|c|}{~}&$N_\tau=11$& 
0.01010(10)&0.004220(35)\\
\multicolumn{1}{|c|}{~}&$N_\tau=12$& 
0.01050(15)&0.004550(35)\\
\multicolumn{1}{|c|}{$\Delta\,\sigma(T)$}&$N_\tau=13$& 
0.01080(20)&0.004750(40)\\
\multicolumn{1}{|c|}{~}&$N_\tau=14$& 
0.01100(25)&0.004910(40)\\
\multicolumn{1}{|c|}{~}&$N_\tau=15$& 
--&0.005020(45)\\
\multicolumn{1}{|c|}{~}&$N_\tau=16$& 
--&0.005110(50)\\
\hline
\multicolumn{2}{|c|}{
$\Delta\,\sigma(0)$}&0.01271(2)&0.00591(1)\\
\hline}
{String tension differences $\Delta\sigma(T)$ as
resulting from  fits to Eq.(\ref{expo}). $\Delta\,\sigma(0)$ is evaluated from 
a fit to Eq.(\ref{fit}) using $\Delta\sigma(T)$ data. 
All the quantities are expressed in lattice spacing units.\label{Table:1}}

Secondly, by flipping a suitable set of couplings, we 
can insert  any Wilson loop or Polyakov correlator directly in the 
Boltzmann factor, producing results with very high precision. 
If, for instance,  we generate a sequence of Monte Carlo configurations where  
the $\sigma_x$ couplings  of all the links crossing the cylindric 
surface bounded by the Polyakov lines $\gamma_{\vec{r}_1}$ 
and $\gamma_{\vec{r}_2}$ are flipped, then the average of whatever 
observable $\Q$ is actually the quantity
\eq
\frac{\bra\,\Q\,P_f(\vec{r}_1)\,P^\dagger_f(\vec{r}_2)\ket_T}{
\bra\,P_f(\vec{r}_1)\,P^\dagger_f(\vec{r}_2)\ket_T}\,.
\en
In our numerical experiment we choose  $\Q= P_f(\vec{r}_1)\,
P^\dagger_f(\vec{r}_2)$, therefore the corresponding averages yield
directly, according to (\ref{pfpff}), the ratio 
\eq
R(\vert\vec{r}_1-\vec{r}_2\vert,T)=\frac{\bra\,P_{ff}(\vec{r}_1)
\,P_{ff}(\vec{r}_2)\ket_T}{
\bra\,P_f(\vec{r}_1)\,P^\dagger_f(\vec{r}_2)\ket_T}\,.
\en
  We estimated this vacuum expectation value with a very powerful 
method, based on the linking properties of the FK clusters \cite{gv}: 
for each FK configuration generated by the above-mentioned algorithm 
one looks for paths in the
 clusters linked with the loops $\gamma_{\vec{r}_1}$ 
and $\gamma_{\vec{r}_2}$. If there is no path of this
 kind we put $ \Q= 1$, otherwise we set $\Q=0$. 
The algorithm we used to determine the linking properties is described 
in~\cite{ziff}.
This method leads to an estimate of $R(\vert\vec{r}_1-\vec{r}_2\vert,T)$ 
with  reduced variance with respect to the conventional numerical 
estimates.

\subsection{Results}
We performed  our Monte Carlo simulations on the AT model, at two different
points of the confining region, for which we  measured previously the 
string tensions \cite{z4kstr_pietro} (see Table 1) . We worked on a  cubic 
periodic lattice of size $128\times128\times N_\tau$ with $N_\tau$ chosen 
in such a way that temperature of our simulations ranged from $T/T_c\simeq0.5$
to $T/T_c\simeq0.8$ and we took the averages over $10^6$ configurations in 
each point.
\FIGURE{\epsfig{file=ckfig/graf-corr-tot.eps,width=10 cm}\caption{Plot of 
$G(r)_{ff}/G(r)_{f}$ on a log scale.
\label{Figure:ratiocorr}}}

The large distance behaviour of the data is well described by  a 
purely exponential behaviour (see Fig.\ref{Figure:ratiocorr})

\FIGURE
{\epsfig{file=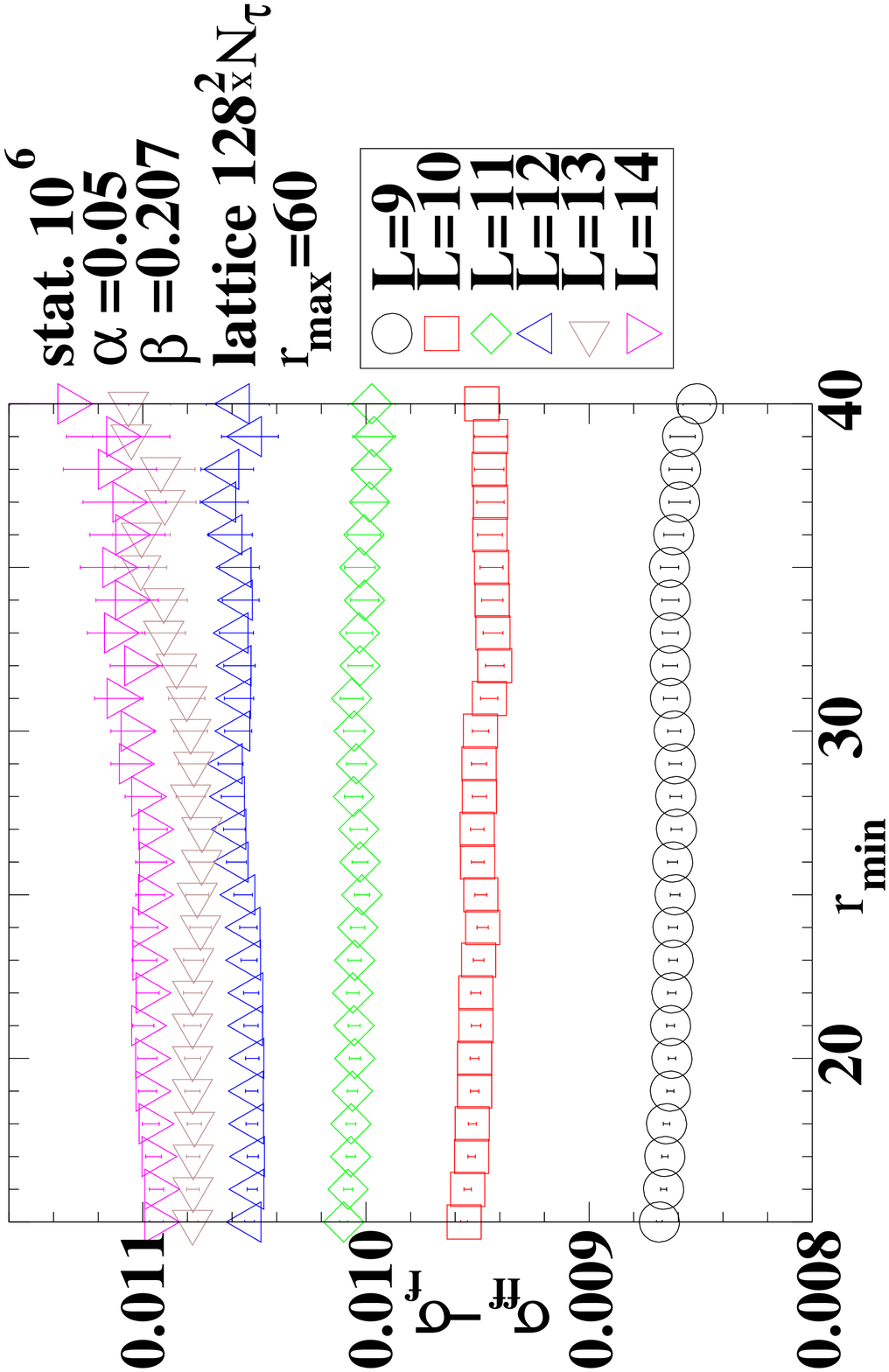,angle=270,width=11 cm}
\caption{The fitted value of $\Delta\sigma$ to Eq.(\ref{expo}) as a function 
of the minimal distance $r_{min}$ considered for different values of $N_\tau$.
The large plateaux show the stability of the fit. 
\label{Figure:4}}}

\eq 
G(r)_{ff}/G(r)_{f}=\frac{\bra\,P_{ff}(\vec{r}_1)\,P_{ff}(\vec{r}_2)\ket_T}
{\bra\,P_f(\vec{r}_1)\,
P^\dagger_f(\vec{r}_2)\ket_T}\propto e^{-\Delta\sigma\,r\,N_\tau}\,,
\label{expo}
\en
with $\Delta\,\sigma=\sigma_{ff}-\sigma_f$. Comparison with (\ref{logppst}) 
shows that the logarithmic term, which is a potential source of systematic 
errors when neglected in Polyakov correlators, here is cancelled in the ratio.
Since (\ref{expo}) is an asymptotic expression, valid in the IR limit, we 
fitted the data to the exponential by progressively discarding the short 
distance points and taking all the values in the range 
$r_{min}\leq r\leq r_{max}=60\,a$ with $r_{min}$ varying from 15 to 40 lattice
spacings $a$. The resulting value of $\Delta\sigma$ turns out to be 
very stable,
as Fig. \ref{Figure:4} shows. The whole set of  values of 
$\Delta\,\sigma(T)$ as functions of the inverse temperature $N_\tau=1/T$ are 
reported in Table 1.

According to Eq.(\ref{main}), in the low temperature limit we expect
the asymptotic behaviour
\eq
\Delta\,\sigma(T)=\Delta\,\sigma(0)\left(1-\frac\pi{6\,\sigma}T^2\right)
     + O(T^3) \,.
\label{fit}
\en
Assuming for $\sigma$ the values estimated in \cite{z4kstr_pietro},
we used $\Delta\sigma(0)$ as the only fitting parameter. Neglecting one or 
two points too close to $T_c$ we got very good fits to (\ref{fit}) 
as shown in Figs. \ref{Figure:5}  and \ref{Figure:6}. 
The fitting parameter $\Delta\,\sigma(0)$ as well as the estimates of
$\Delta\,\sigma(T)$ are reported in Table 1. Note that $\Delta\,\sigma(0)$ 
agrees with the difference of the previous estimates  \cite{z4kstr_pietro}, 
also reported in Table 1, however the error is reduced by a factor of 25 
(first set of data) and even of 50 (second set of data).
The reason of this gain in precision is due to the fact that 
$\sigma_2$ was evaluated from a fit to (\ref{ppst}), even taking in account the 
Next-to-Leading-Order terms, which was rather poor 
because the 2-string does not behave as a free bosonic string 
\cite{z4kstr_pietro}. On the contrary our fits to (\ref{expo}) and 
(\ref{fit}) are very stable and the corresponding reduced $\chi^2/d.o.f$ 
are of the order of 1 or less.   
\FIGURE
{\epsfig{file=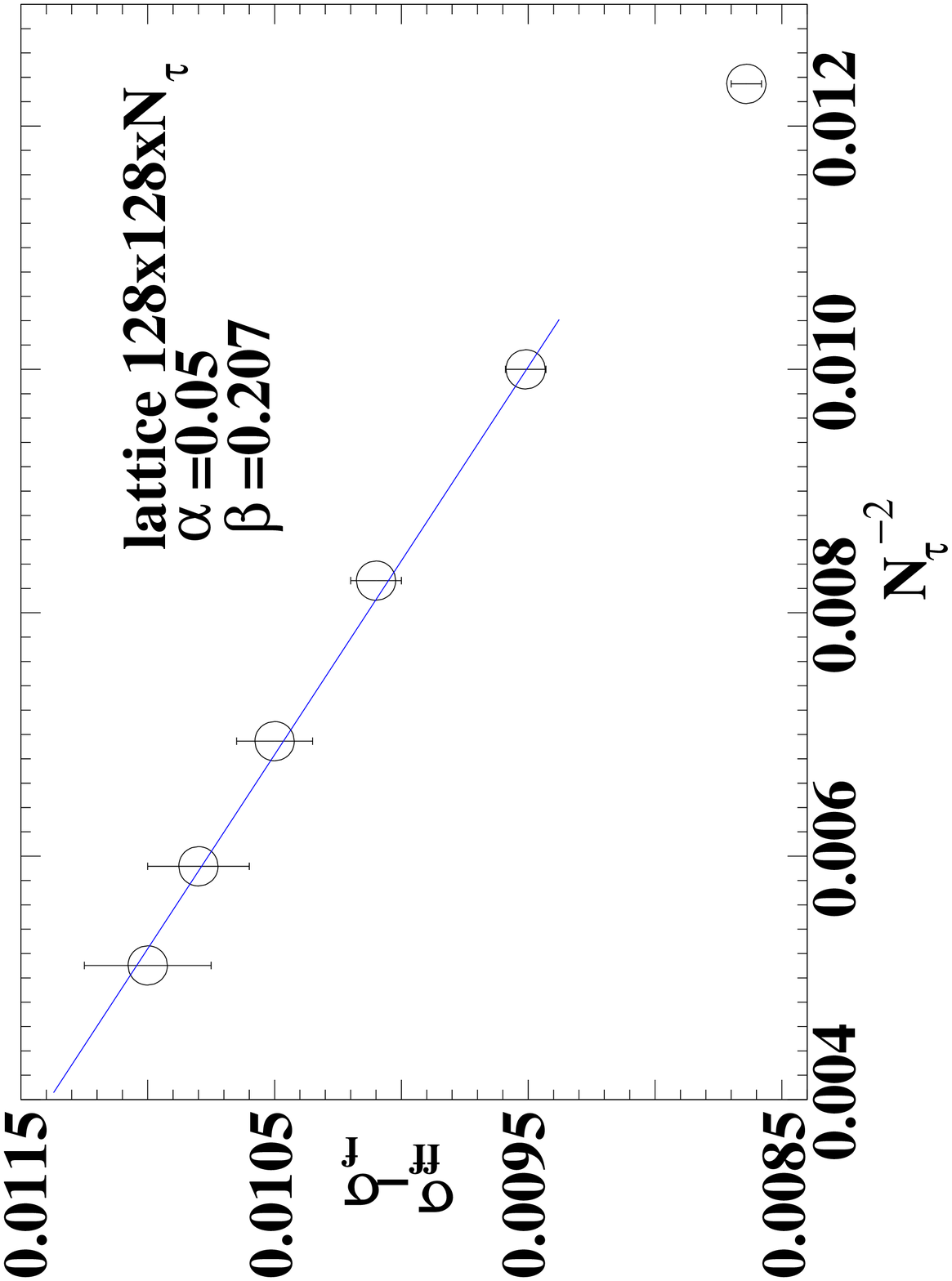,angle=270,width=10 cm}
\caption{Values of $\Delta\sigma(T)$  from the first set of data of Table 1 
versus $T^2=1/N^2_\tau$. The solid line is 
a one-parameter fit to Eq.(\ref{fit}).
\label{Figure:5}}}
\FIGURE
{\epsfig{file=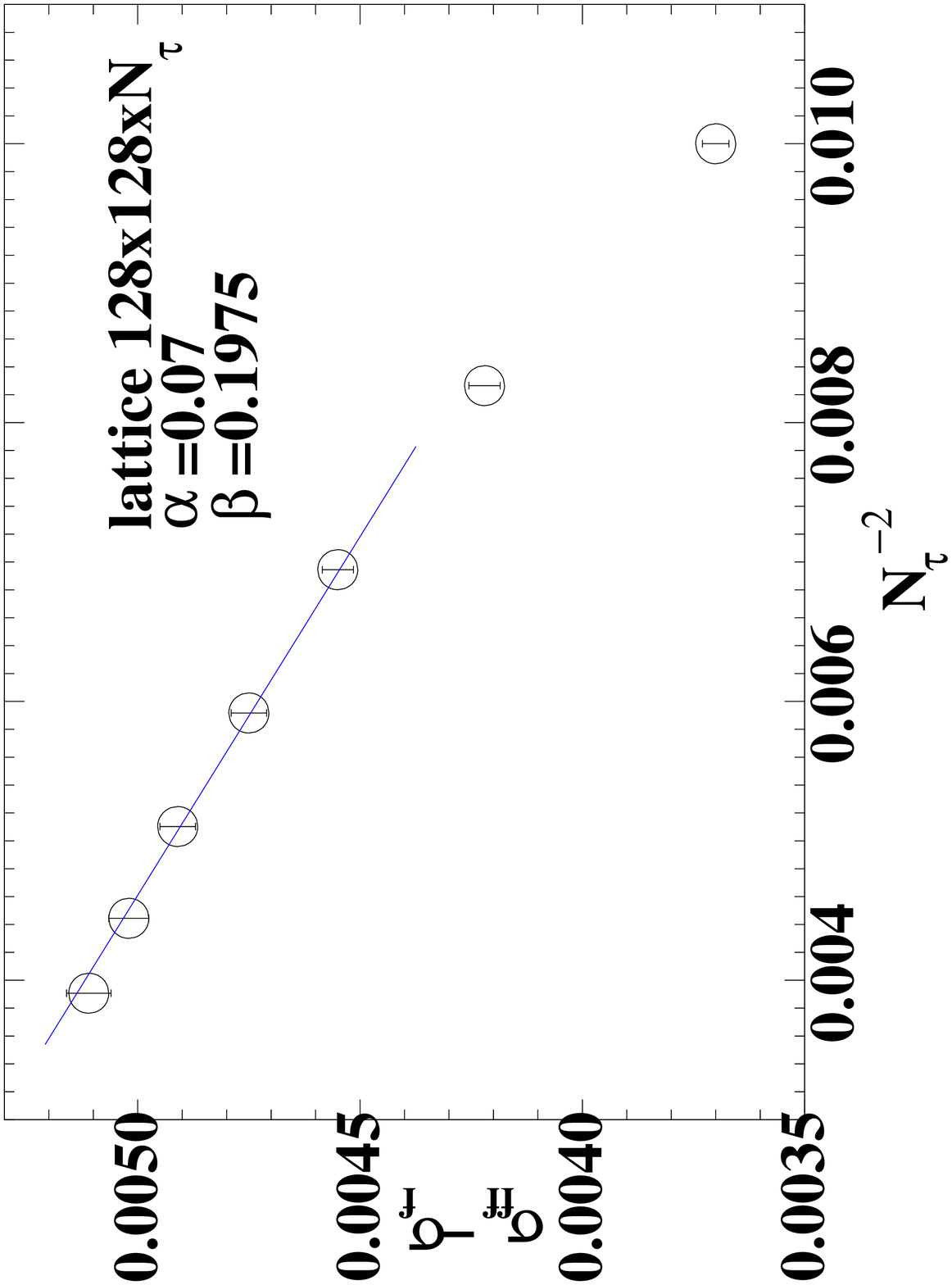,angle=270,width=10 cm}
\caption{Same as Fig.\ref{Figure:5}, but for the second set of data of Table 1.
\label{Figure:6}}}
\section{Discussion}
Our calculations in this paper were in two parts. In the first part we investigated  the string nature of long flux tubes generated in an $\sun$ gauge theory 
by external sources in representations with $N-$ality $k>1$. In a previous 
work \cite{z4kstr_pietro} we were led to the conclusion that these k-strings 
are not adequately described by the free bosonic string picture. In this paper 
we argued that the central charge of the effective k-string theory is not 
simply $d-2$, like in  the free bosonic string, but $c_k=(d-2)\frac{\sigma_k}\sigma$. This simple 
recipe was a straightforward consequence of demanding that the asymptotic 
functional form of Polyakov correlators associated to the baryon vertex 
should not change when varying the mutual positions of the Polyakov lines, 
the reason being that certain positions create k-strings inside the baryon 
vertex which modify the functional form of the correlator unless $\frac{\sigma_k(T)}{\sigma(T)}={\rm constant}+O(T^3)$. This in turn fixes unambiguously the value of $c_k$. Although the geometric derivation is quite general and the resulting expression for $c_k$ is appealing for its simplicity, it would be very 
important to find some independent quantum argument in support of it.

The second part of the paper dealt with 2-strings in a 3D $\Z_4$ gauge model 
and in particular with the difference $\Delta\sigma(T)=\sigma_2(T)-\sigma(T)$ of the string 
tensions as a function of the temperature. Combining together three essential
ingredients, namely the duality transformation, an efficient non-local 
cluster algorithm and finally a choice of flipped links which allows to 
directly measure  
the ratio of Polyakov correlators  belonging to different representations, 
we obtained values for $ \Delta\sigma(T)$ with unprecedented precision which nicely agree with our general formula (\ref{main}).
There is however much scope for improving these calculations: as Fig. 
\ref{Figure:5} and  \ref{Figure:6} show, with little more effort it would be possible to evaluate also the corrections of order $T^3$ and $T^4$, that in the
case of fundamental string are expected to be universal.
\acknowledgments
We are grateful to Michele Caselle, Paolo Grinza and Ettore Vicari for useful
discussions.

\end{document}